\documentclass[preprint,showpacs,showkeys,amsmath,amssymb]{revtex4}

\usepackage{graphicx}

\allowdisplaybreaks

\begin{document}

\title{Virtual Photon Correction to the $K^+\rightarrow\pi^+\pi^0\pi^0$ Decay}

\author{A.~Nehme}
\email{miryama.nehme@wanadoo.fr}
\affiliation{
27 rue du Four de la Terre \\
F-84000 Avignon, France}

\date{\today}

\begin{abstract}
We consider electromagnetic corrections to the non-leptonic kaon
decay, $K^+\rightarrow\pi^+\pi^0\pi^0$, due to explicit virtual
photons only. The decay amplitude is calculated at one-loop level in
the framework of Chiral Perturbation Theory. The interest in this
process is twofold: It is actually measured by the NA48
collaboration from one side, and, the value of the amplitude at the
$\pi\pi$ threshold gives access to $\pi\pi$ scattering lengths from
the other side. We found that the present correction is about $5$ to
$6\%$ the value of the Born amplitude squared. Combined with another
piece published recently, this fixes the size of isospin breaking
correction to the amplitude squared to $7\%$ its one-loop level
value in the absence of isospin breaking and at the center of Dalitz
plot.
\end{abstract}


\maketitle

\section{Introduction}

Starting May, Nicola Cabibbo submitted to the arxiv a letter
proposing a ``potentially accurate'' method for the determination of
$\pi\pi$ scattering length $a_0^0-a_0^2$ from the $\pi^0\pi^0$
spectrum in $K^+\rightarrow\pi^+\pi^0\pi^0$ near the $\pi^+\pi^-$
threshold~\cite{Cabibbo:2004gq}. Next day, Hans Bijnens and Fredrik
Borg submitted a paper treating isospin breaking in $K\rightarrow
3\pi$ decays in the framework of chiral perturbation
theory~\cite{Bijnens:2004ku}. In their study, the authors of
Ref.~\cite{Bijnens:2004ku} took into account strong isospin breaking
and local electromagnetic corrections. The former is proportional to
the up and down quark mass difference $m_u-m_d$. With respect to
local electromagnetic correction, it includes the effects of mass
square difference between charged and neutral pions
$M_{\pi^{\pm}}^2-M_{\pi^0}^2$ and electroweak counterterms.

Let us justify the necessity of the present work. The NA48
collaboration investigates actually the $CP$ violating asymmetry in
$K^{\pm}\rightarrow\pi^{\pm}\pi^0\pi^0$ decays. So, there is no need
in principle to upgrade the experimental setup in order to study the
$\pi^0\pi^0$ spectrum near the $\pi^+\pi^-$ threshold. From the
theoretical point of view, the $\pi\pi$ scattering length
$a_0^0-a_0^2$ is very sensitive to the value of the quark
condensate. Hence, an accurate measurement of this quantity would
allow a best understanding of the quantum chromodynamics vacuum
structure.

The most convenient theoretical framework to study non-leptonic weak
kaon decays is chiral perturbation theory extended to include
non-leptonic weak interactions~\cite{Kambor:1989tz}. The effective
Lagrangian constructed in Ref.~\cite{Kambor:1989tz} has been used
in~\cite{Kambor:ah, Bijnens:2002vr, Gamiz:2003pi} to calculate
$K\rightarrow 3\pi$ decays to next-to-leading chiral order. The
theoretical prediction did not match the experimental data.
Naturally, the need to push the chiral expansion up to higher orders
is imperative. In a parallel direction, one has to improve the
theoretical prediction by evaluating the size of isospin breaking
corrections. To do so, electromagnetism should be incorporated
appropriately in the effective Lagrangian
formalism~\cite{Ecker:2000zr}.

We shall distinguish between two types of virtual photons: hard and
soft. Hard photons are integrated out from the effective theory
leaving local electromagnetic interactions weighted by order
parameters. The latter are counterterms in the sense that they
absorb ultraviolet divergences produced by soft virtual photon
loops. The isospin breaking corrections calculated in
Ref.~\cite{Bijnens:2004ku} are due to $m_u-m_d$ and to hard virtual
photons. Obviously, they are ultraviolet divergent. Moreover, their
size is about $10\%$ the value of the one-loop level squared
amplitude in the absence of isospin breaking. The authors of
Ref.~\cite{Bijnens:2004ku} concluded that the correction they
obtained is not ``quite enough to solve the discrepancies'' between
theory and experiment and that the effects of soft virtual photons
should be included. The aim of the present work is to evaluate these
effects.

\section{Kinematics}

The decay process
\begin{equation} \label{eq:process}
K^+(k)\rightarrow\pi^0(p_1)\pi^0(p_2)\pi^+(p_3)
\end{equation}
is studied in terms of three scalars
\begin{equation} \label{eq:mandelstam}
s_i\,\doteq\,(k-p_i)^2\,, \qquad i\,=\,1,\,2,\,3\,,
\end{equation}
satisfying the on-shell condition
\begin{equation} \label{eq:onshell}
\sum_is_i\,=\,M_{K^{\pm}}^2+M_{\pi^{\pm}}^2+2M_{\pi^0}^2\,.
\end{equation}
The differential decay rate is defined by
\begin{equation} \label{eq:rate}
d\Gamma\,\doteq\,\frac{1}{4M_{K^{\pm}}}\,\mid\mathcal{M}\mid^2d\Phi\,,
\end{equation}
with the differential phase space
\begin{equation}
d\Phi\,\doteq\,(2\pi
)^4\delta^{4}\left(\sum_ip_i-k\right)\prod_i\frac{d^3p_i}{(2\pi
)^32p_i^0}\,,
\end{equation}
and the decay amplitude $\mathcal{M}$. The spectrum of the phase
space with respect to $s_3$ reads
\begin{equation}
\frac{d\Phi}{ds_3}\,=\,\frac{1}{2^7\pi^3}\,\frac{\lambda^{1/2}(s_3,M_{\pi^0}^2,M_{\pi^0}^2)}{s_3}\,\frac{\lambda^{1/2}(s_3,M_{\pi^{\pm}}^2,M_{K^{\pm}}^2)}{M_{K^{\pm}}^2}\,,
\end{equation}
where $\lambda$ stands for the K\"{a}ll\'{e}n function
\begin{equation}
\lambda (x,y,z)\,\doteq\,x^2+y^2+z^2-2xy-2xz-2yz\,.
\end{equation}
The physical region is bounded by
\begin{equation}
(M_{\pi^{\pm}}+M_{\pi^0})^2\,\leqslant\,s_1\,,\,s_2\,\leqslant\,(M_{K^{\pm}}-M_{\pi^0})^2\,,
\quad
4M_{\pi^0}^2\,\leqslant\,s_3\,\leqslant\,(M_{K^{\pm}}-M_{\pi^{\pm}})^2\,.
\label{eq:physical-region}
\end{equation}
In what follows, we set $M_{K^{\pm}}=M_K$ and
$M_{\pi^{\pm}}=M_{\pi^0}=M_{\pi}$. This is safe since
$e^2M_{\pi^{\pm}}^2=e^2M_{\pi}^2+\mathcal{O}(e^4)$.

\section{The decay amplitude}

The decay (\ref{eq:process}) can proceed locally. It can also be
mediated by pion or kaon multi-poles. Let us then write the decay
amplitude at one-loop level as~\footnote{For comparison, the
amplitude in Ref.~\cite{Bijnens:2004ku} reads
$A_{00+}=i\mathcal{M}$.}:
\begin{equation} \label{eq:amplitude}
\mathcal{M}\,\doteq\,\mathcal{T}\,+\,\frac{\dot{\mathcal{Z}}}{M_K^2-M_{\pi}^2}\,+\,\frac{\ddot{\mathcal{Z}}}{(M_K^2-M_{\pi}^2)^2}\,.
\end{equation}
We follow notations of Ref.~\cite{Bijnens:2004ku} for the
strangness-changing Lagrangian parameters. The decay amplitude takes
then the following form:
\begin{equation}
X\,=\,CG_8X^{(8)}+CG_8'X^{(8')}+CG_{27}X^{(27)}\,, \qquad
X\,=\,\mathcal{M}\,,\,\mathcal{T}\,,\,\dot{\mathcal{Z}}\,,\,\ddot{\mathcal{Z}}\,.
\end{equation}
Moreover, the subscripts $0$ and $1$ will be assigned to Born
amplitudes and one-loop explicit virtual photon corrections,
respectively.

\subsection{Born amplitude}

The lowest order in the perturbative expansion of the amplitude
comprises five diagrams as sketched in Fig.~$1$. We present the
result in terms of scalar products avoiding any ambiguity
\begin{eqnarray}
\mathcal{T}_0^{(8)} &=& \frac{1}{3}\,(k\cdot p_1+k\cdot p_2-4k\cdot
p_3-p_1\cdot p_3-p_2\cdot p_3)\,,\\
\mathcal{T}_0^{(8')} &=& \frac{1}{3}\,M_K^2\,, \\
\mathcal{T}_0^{(27)} &=& \frac{1}{9}\,(2k\cdot p_1+2k\cdot
p_2-8k\cdot
p_3-30p_1\cdot p_2+13p_1\cdot p_3+13p_2\cdot p_3)\,,\\
\dot{\mathcal{Z}}_0^{(8)} &=& -\frac{1}{3}\,(M_K^2-M_{\pi}^2)\times
\nonumber\\
&& (M_{\pi}^2+2k\cdot p_1+2k\cdot p_2-4k\cdot p_3+4p_1\cdot
p_2-2p_1\cdot p_3-2p_2\cdot
p_3)\,, \\
\dot{\mathcal{Z}}_0^{(8')} &=&
-\frac{1}{3}\,(M_K^2-M_{\pi}^2)M_K^2\,, \\
\dot{\mathcal{Z}}_0^{(27)} &=&
-\frac{1}{18}\left[4M_{\pi}^2(M_K^2-M_{\pi}^2)\right.
\nonumber\\
&& \left. +(8M_K^2+7M_{\pi}^2)(k\cdot p_1+k\cdot p_2-2k\cdot
p_3+2p_1\cdot p_2-p_1\cdot p_3-p_2\cdot p_3)\right]\,.
\end{eqnarray}
Note that $\mathcal{M}_0^{(8')}$ vanishes if we add all
contributions.

\subsection{Virtual photons}

In this section, we shall calculate one-loop diagrams with a virtual
photon exchanged between two meson legs, one meson leg and a vertex,
and two vertices. The various topologies are depicted in Figs.~($2$)
- ($9$). The loop functions figuring in the expressions are
tabulated in the appendix. A fictitious mass $m_{\gamma}$ is
attributed to the photon in order to regularize infrared divergent
integrals. For completeness, we calculate the amplitudes
proportional to $G_8'$. Only those proportional to $G_8$ and
$G_{27}$ will be taken into account numerically. We shall give the
contribution of each figure separately.

\subsubsection*{Figure~$2$}
We have
\begin{eqnarray}
\mathcal{T}_1^{(8)} &=& \frac{e^2}{3}\,\left\lbrace
2A_0(M_{\pi})+2A_0(M_K)\right.
\nonumber\\
&& -2(4\pi )^{-2}(k\cdot p_1+k\cdot p_2-4k\cdot p_3-p_1\cdot
p_3-p_2\cdot p_3)[1+\ln (m_{\gamma}^2)-\ln (M_{\pi}M_K)]
\nonumber\\
&& +(8M_{\pi}^2+k\cdot p_1+k\cdot p_2-4k\cdot p_3-p_1\cdot
p_3-p_2\cdot p_3)B_0(-p_3,0,M_{\pi})
\nonumber\\
&& +2(4M_{\pi}^2-k\cdot p_1-k\cdot p_2+4k\cdot p_3+p_1\cdot
p_3+p_2\cdot p_3)B_1(-p_3,0,M_{\pi})
\nonumber\\
&& +(8M_K^2+k\cdot p_1+k\cdot p_2-4k\cdot p_3-p_1\cdot p_3-p_2\cdot
p_3)B_0(-k,0,M_K)
\nonumber\\
&& +2(4M_K^2-k\cdot p_1-k\cdot p_2+4k\cdot p_3+p_1\cdot p_3+p_2\cdot
p_3)B_1(-k,0,M_K)
\nonumber\\
&& -(k\cdot p_1+k\cdot p_2-4k\cdot p_3-p_1\cdot p_3-p_2\cdot
p_3)B_0(p_3-k,M_{\pi},M_K)
\nonumber\\
&& +4(k\cdot p_3)(k\cdot p_1+k\cdot p_2-4k\cdot p_3
\nonumber\\
&& \left. -p_1\cdot p_3-p_2\cdot p_3)C_0(-p_3,-k,m_{\gamma},M_{\pi},M_K)\right\rbrace\,, \\
\mathcal{T}_1^{(8')} &=& \frac{e^2}{3}\,M_K^2\left\lbrace -2(4\pi
)^{-2}[1+\ln (m_{\gamma}^2)-\ln (M_{\pi}M_K)]\right.
\nonumber\\
&&
+B_0(-p_3,0,M_{\pi})-2B_1(-p_3,0,M_{\pi})+B_0(-k,0,M_K)-2B_1(-k,0,M_K)
\nonumber\\
&& \left. -B_0(p_3-k,M_{\pi},M_K)+4(k\cdot
p_3)C_0(-p_3,-k,m_{\gamma},M_{\pi},M_K)\right\rbrace\,, \\
\mathcal{T}_1^{(27)} &=& \frac{e^2}{9}\left\lbrace
4A_0(M_{\pi})+4A_0(M_K)\right.
\nonumber\\
&& -2(4\pi )^{-2}(2k\cdot p_1+2k\cdot p_2-8k\cdot p_3
\nonumber\\
&& -30p_1\cdot p_2+13p_1\cdot p_3+13p_2\cdot p_3)[1+\ln
(m_{\gamma}^2)-\ln (M_{\pi}M_K)]
\nonumber\\
&& +(16M_{\pi}^2+2k\cdot p_1+2k\cdot p_2-8k\cdot p_3
\nonumber\\
&& -30p_1\cdot p_2-17p_1\cdot p_3-17p_2\cdot p_3)B_0(-p_3,0,M_{\pi})
\nonumber\\
&& +2(8M_{\pi}^2-2k\cdot p_1-2k\cdot p_2+8k\cdot p_3
\nonumber\\
&& +30p_1\cdot p_2-13p_1\cdot p_3-13p_2\cdot p_3)B_1(-p_3,0,M_{\pi})
\nonumber\\
&& +(16M_K^2-28k\cdot p_1-28k\cdot p_2-8k\cdot p_3
\nonumber\\
&& -30p_1\cdot p_2+13p_1\cdot p_3+13p_2\cdot p_3)B_0(-k,0,M_K)
\nonumber\\
&& +2(8M_K^2-2k\cdot p_1-2k\cdot p_2+8k\cdot p_3
\nonumber\\
&& +30p_1\cdot p_2-13p_1\cdot p_3-13p_2\cdot p_3)B_1(-k,0,M_K)
\nonumber\\
&& -2(k\cdot p_1+k\cdot p_2-4k\cdot p_3-15p_1\cdot p_2-p_1\cdot
p_3-p_2\cdot p_3)B_0(p_3-k,M_{\pi},M_K)
\nonumber\\
&& -15(k\cdot p_1+k\cdot p_2-p_1\cdot p_3-p_2\cdot
p_3)B_1(p_3-k,M_{\pi},M_K)
\nonumber\\
&& +4(k\cdot p_3)(2k\cdot p_1+2k\cdot p_2-8k\cdot p_3
\nonumber\\
&& -30p_1\cdot p_2+13p_1\cdot p_3+13p_2\cdot
p_3)C_0(-p_3,-k,m_{\gamma},M_{\pi},M_K)
\nonumber\\
&& +60(k\cdot p_3)(p_1\cdot p_3+p_2\cdot
p_3)C_1(-p_3,-k,0,M_{\pi},M_K)
\nonumber\\
&& \left. +60(k\cdot p_3)(k\cdot p_1+k\cdot
p_2)C_2(-p_3,-k,0,M_{\pi},M_K)\right\rbrace \,.
\end{eqnarray}

\subsubsection*{Figure~$3$}
We have
\begin{eqnarray}
\mathcal{T}_1^{(8)} &=& \frac{3}{2}\,\mathcal{T}_1^{(27)} \\
&=& -\frac{2e^2}{3}\left\lbrace
A_0(M_{\pi})-A_0(M_K)-M_K^2B_0(-p_3,0,M_{\pi})\right.
\nonumber\\
&& -M_{\pi}^2B_0(0,M_{\pi},M_K) +M_{\pi}^2B_0(k-p_3,M_{\pi},M_{\pi})
\nonumber\\
&& +(4M_K^2+3M_{\pi}^2+k\cdot p_1+k\cdot p_2-2k\cdot p_3
\nonumber\\
&& +2p_1\cdot p_2-p_1\cdot p_3-p_2\cdot p_3)B_0(-k,0,M_K)
\nonumber\\
&& -(3M_K^2+3M_{\pi}^2+k\cdot p_1+k\cdot p_2-2k\cdot p_3
\nonumber\\
&& +2p_1\cdot p_2-p_1\cdot p_3-p_2\cdot p_3)B_0(-k,0,M_{\pi})
\nonumber\\
&& -(M_{\pi}^2+k\cdot p_1+k\cdot p_2-2k\cdot p_3
\nonumber\\
&& +2p_1\cdot p_2-p_1\cdot p_3-p_2\cdot p_3)B_0(p_3-k,M_{\pi},M_K)
\nonumber\\
&& +(M_K^2+k\cdot p_1+k\cdot p_2-2k\cdot p_3
\nonumber\\
&& +2p_1\cdot p_2-p_1\cdot p_3-p_2\cdot
p_3)B_0(p_3-k,M_{\pi},M_{\pi})
\nonumber\\
&& +(M_{\pi}^2+4k\cdot p_3)(M_{\pi}^2+k\cdot p_1+k\cdot p_2-2k\cdot
p_3
\nonumber\\
&& +2p_1\cdot p_2-p_1\cdot p_3-p_2\cdot
p_3)C_0(-p_3,-k,m_{\gamma},M_{\pi},M_K)
\nonumber\\
&& +M_{\pi}^2(2M_K^2+3M_{\pi}^2+k\cdot p_1+k\cdot p_2-2k\cdot p_3
\nonumber\\
&& +2p_1\cdot p_2-p_1\cdot p_3-p_2\cdot p_3)C_0(-k,-k,0,M_{\pi},M_K)
\nonumber\\
&& +(M_K^2-M_{\pi}^2-4k\cdot p_3)(M_K^2+M_{\pi}^2+k\cdot p_1+k\cdot
p_2-2k\cdot p_3
\nonumber\\
&& +2p_1\cdot p_2-p_1\cdot p_3-p_2\cdot
p_3)C_0(-p_3,-k,0,M_{\pi},M_{\pi})
\nonumber\\
&& -M_{\pi}^2(M_{\pi}^2+k\cdot p_1+k\cdot p_2-2k\cdot p_3
\nonumber\\
&& +2p_1\cdot p_2-p_1\cdot p_3-p_2\cdot
p_3)C_0(0,k-p_3,M_K,M_{\pi},M_{\pi})
\nonumber\\
&& -M_{\pi}^2(M_K^2-M_{\pi}^2-4k\cdot p_3)(M_{\pi}^2+k\cdot
p_1+k\cdot p_2-2k\cdot p_3
\nonumber\\
&& \left. +2p_1\cdot p_2-p_1\cdot p_3-p_2\cdot p_3)D_0(-k,-p_3,-k,m_{\gamma},M_K,M_{\pi},M_{\pi})\right\rbrace \,, \\
\mathcal{T}_1^{(8')} &=& \frac{2e^2}{3}\,M_K^2\left\lbrace
B_0(k-p_3,M_{\pi},M_{\pi})-B_0(0,M_{\pi},M_K)\right.
\nonumber\\
&& -B_0(-k,0,M_{\pi})+2B_0(-k,0,M_K)-B_0(-p_3,0,M_{\pi})
\nonumber\\
&& +(M_{\pi}^2+k\cdot p_1+k\cdot p_2-2k\cdot p_3
\nonumber\\
&& +2p_1\cdot p_2-p_1\cdot p_3-p_2\cdot
p_3)C_0(-p_3,-k,m_{\gamma},M_{\pi},M_K)
\nonumber\\
&& +(2M_K^2+3M_{\pi}^2+k\cdot p_1+k\cdot p_2-2k\cdot p_3
\nonumber\\
&& +2p_1\cdot p_2-p_1\cdot p_3-p_2\cdot p_3)C_0(-k,-k,0,M_{\pi},M_K)
\nonumber\\
&& +(M_K^2-M_{\pi}^2-4k\cdot p_3)C_0(-p_3,-k,0,M_{\pi},M_{\pi})
\nonumber\\
&& -(M_{\pi}^2+k\cdot p_1+k\cdot p_2-2k\cdot p_3
\nonumber\\
&& +2p_1\cdot p_2-p_1\cdot p_3-p_2\cdot
p_3)C_0(0,k-p_3,M_K,M_{\pi},M_{\pi})
\nonumber\\
&& -(M_K^2-M_{\pi}^2-4k\cdot p_3)(M_{\pi}^2+k\cdot p_1+k\cdot
p_2-2k\cdot p_3
\nonumber\\
&& \left. +2p_1\cdot p_2-p_1\cdot p_3-p_2\cdot
p_3)D_0(-k,-p_3,-k,m_{\gamma},M_K,M_{\pi},M_{\pi})\right\rbrace \,.
\end{eqnarray}

\subsubsection*{Figure~$4$}
We have
\begin{eqnarray}
\dot{\mathcal{Z}}_1^{(8)} &=&
-\dot{\mathcal{Z}}_1^{(8')}\,=\,\frac{3}{2}\,\dot{\mathcal{Z}}_1^{(27)} \\
&=& -\frac{2e^2}{3}\,M_K^2\left\lbrace -2A_0(M_{\pi})\right.
\nonumber\\
&& +(M_K^2+4M_{\pi}^2+k\cdot p_1+k\cdot p_2-2k\cdot p_3
\nonumber\\
&& +2p_1\cdot p_2-p_1\cdot p_3-p_2\cdot p_3)B_0(-p_3,0,M_{\pi})
\nonumber\\
&& +(3M_K^2+2M_{\pi}^2+k\cdot p_1+k\cdot p_2-2k\cdot p_3
\nonumber\\
&& +2p_1\cdot p_2-p_1\cdot p_3-p_2\cdot p_3)B_0(-k,0,M_{\pi})
\nonumber\\
&& -(M_K^2+k\cdot p_1+k\cdot p_2-2k\cdot p_3
\nonumber\\
&& +2p_1\cdot p_2-p_1\cdot p_3-p_2\cdot
p_3)B_0(p_3-k,M_{\pi},M_{\pi})
\nonumber\\
&& -(M_K^2-M_{\pi}^2-4k\cdot p_3)(M_K^2+k\cdot p_1+k\cdot
p_2-2k\cdot p_3
\nonumber\\
&& \left. +2p_1\cdot p_2-p_1\cdot p_3-p_2\cdot
p_3)C_0(-p_3,-k,0,M_{\pi},M_{\pi})\right\rbrace \,.
\end{eqnarray}

\subsubsection*{Figure~$5$}
We have
\begin{eqnarray}
\dot{\mathcal{Z}}_1^{(8)} &=&
\frac{3}{2}\,\dot{\mathcal{Z}}_1^{(27)} \\
&=& -\frac{2e^2}{3}\,(M_{\pi}^2+k\cdot p_1+k\cdot p_2-2k\cdot
p_3+2p_1\cdot p_2-p_1\cdot p_3-p_2\cdot p_3)\times
\nonumber\\
&& \left\lbrace -2M_K^2(4\pi )^{-2}[1+\ln (m_{\gamma}^2)-\ln
(M_{\pi}M_K)]\right.
\nonumber\\
&& +A_0(M_{\pi})-M_{\pi}^2B_0(0,M_{\pi},M_K)
\nonumber\\
&& +M_{\pi}^2B_0(-k,0,M_K)-2M_K^2B_1(-k,0,M_K)
\nonumber\\
&& -2M_K^2B_1(-p_3,0,M_{\pi})-(2M_K^2+M_{\pi}^2)B_0(-k,0,M_{\pi})
\nonumber\\
&& \left. +M_{\pi}^2(3M_K^2+M_{\pi}^2)C_0(-k,-k,0,M_{\pi},M_K)\right\rbrace \,, \\
\dot{\mathcal{Z}}_1^{(8')} &=&
\frac{2e^2}{3}\,M_K^2(M_{\pi}^2+k\cdot p_1+k\cdot p_2-2k\cdot
p_3+2p_1\cdot p_2-p_1\cdot p_3-p_2\cdot p_3)\times
\nonumber\\
&& \left\lbrace -2(4\pi )^{-2}[1+\ln (m_{\gamma}^2)-\ln
(M_{\pi}M_K)]\right.
\nonumber\\
&& -2B_1(-p_3,0,M_{\pi})-B_0(0,M_{\pi},M_K)+B_0(-k,0,M_{\pi})
\nonumber\\
&& +B_0(-k,0,M_K)-2B_1(-k,0,M_K)
\nonumber\\
&& \left.
+(3M_K^2+M_{\pi}^2)C_0(-k,-k,0,M_{\pi},M_K)\right\rbrace\,, \\
\ddot{\mathcal{Z}}_1^{(8)} &=&
-\ddot{\mathcal{Z}}_1^{(8')}\,=\,\frac{3}{2}\,\ddot{\mathcal{Z}}_1^{(27)}
\\
&=& -\frac{2e^2}{3}\,M_K^2(M_{\pi}^2+k\cdot p_1+k\cdot p_2-2k\cdot
p_3+2p_1\cdot p_2-p_1\cdot p_3-p_2\cdot p_3)\times \nonumber\\
&& \left[
-A_0(M_{\pi})+2(M_K^2+M_{\pi}^2)B_0(-k,0,M_{\pi})\right]\,.
\end{eqnarray}

\subsubsection*{Figure~$6$}
We have
\begin{eqnarray}
\mathcal{T}_1^{(8)} &=& \frac{3}{2}\,\mathcal{T}_1^{(27)} \\
&=&
-\frac{e^2}{6}\left\{-A_0(M_{\pi})+A_0(M_K)-M_{\pi}^2B_0(-k,0,M_K)\right.
\nonumber\\
&& +(M_K^2+6M_{\pi}^2+k\cdot p_1+k\cdot p_2-2k\cdot p_3
\nonumber\\
&& +2p_1\cdot
p_2-p_1\cdot p_3-p_2\cdot p_3)B_0(-p_3,0,M_{\pi}) \nonumber\\
&& -(M_K^2+5M_{\pi}^2+k\cdot p_1+k\cdot p_2-2k\cdot p_3 \nonumber\\
&& +2p_1\cdot
p_2-p_1\cdot p_3-p_2\cdot p_3)B_0(-p_3,0,M_K) \nonumber\\
&& +M_K^2B_0(p_3-k,M_K,M_K)-M_K^2B_0(0,M_{\pi},M_K) \nonumber\\
&& -(2M_K^2+2M_{\pi}^2+k\cdot p_1+k\cdot p_2-2k\cdot p_3
\nonumber\\
&& +2p_1\cdot p_2-p_1\cdot p_3-p_2\cdot p_3)B_0(p_3-k,M_{\pi},M_K)
\nonumber\\
&& +(M_K^2+3M_{\pi}^2+k\cdot p_1+k\cdot p_2-2k\cdot p_3\nonumber\\
&& +2p_1\cdot p_2-p_1\cdot p_3-p_2\cdot p_3)B_0(k-p_3,M_K,M_K)
\nonumber\\
&& +(M_K^2+4k\cdot p_3)(2M_K^2+2M_{\pi}^2+k\cdot p_1+k\cdot
p_2-2k\cdot p_3 \nonumber\\
&& +2p_1\cdot p_2-p_1\cdot p_3-p_2\cdot
p_3)C_0(-p_3,-k,m_{\gamma},M_{\pi},M_K) \nonumber\\
&& +M_K^2(M_K^2+7M_{\pi}^2+k\cdot p_1+k\cdot p_2-2k\cdot p_3
\nonumber\\
&& +2p_1\cdot p_2-p_1\cdot p_3-p_2\cdot
p_3)C_0(-p_3,-p_3,0,M_{\pi},M_K) \nonumber\\
&& -(M_K^2-M_{\pi}^2+4k\cdot p_3)(2M_K^2+3M_{\pi}^2+k\cdot
p_1+k\cdot p_2-2k\cdot p_3 \nonumber\\
&& +2p_1\cdot p_2-p_1\cdot p_3-p_2\cdot p_3)C_0(-p_3,-k,0,M_K,M_K)
\nonumber\\
&& -M_K^2(2M_K^2+2M_{\pi}^2+k\cdot p_1+k\cdot p_2-2k\cdot p_3
\nonumber\\
&& +2p_1\cdot p_2-p_1\cdot p_3-p_2\cdot
p_3)C_0(0,p_3-k,M_{\pi},M_K,M_K) \nonumber\\
&& +M_K^2(M_K^2-M_{\pi}^2+4k\cdot p_3)(2M_K^2+2M_{\pi}^2+k\cdot
p_1+k\cdot p_2-2k\cdot p_3 \nonumber\\
&& \left. +2p_1\cdot p_2-p_1\cdot p_3-p_2\cdot
p_3)D_0(-p_3,-k,-p_3,m_{\gamma},M_{\pi},M_K,M_K)\right\} \,, \\
\mathcal{T}_1^{(8')} &=&
\frac{e^2}{6}\,M_K^2\left\{-B_0(-k,0,M_K)-B_0(-p_3,0,M_{\pi})
\right. \nonumber\\
&& +B_0(p_3-k,M_K,M_K)+2B_0(-p_3,0,M_K)-B_0(0,M_{\pi},M_K)
\nonumber\\
&& +(2M_K^2+2M_{\pi}^2+k\cdot p_1+k\cdot p_2-2k\cdot p_3 \nonumber\\
&& +2p_1\cdot p_2-p_1\cdot p_3-p_2\cdot
p_3)C_0(-p_3,-k,m_{\gamma},M_{\pi},M_K) \nonumber\\
&& +(M_K^2+7M_{\pi}^2+k\cdot p_1+k\cdot p_2-2k\cdot p_3 \nonumber\\
&& +2p_1\cdot p_2-p_1\cdot p_3-p_2\cdot
p_3)C_0(-p_3,-p_3,0,M_{\pi},M_K) \nonumber\\
&& -(2M_K^2+2M_{\pi}^2+k\cdot p_1+k\cdot p_2-2k\cdot p_3 \nonumber\\
&& +2p_1\cdot p_2-p_1\cdot p_3-p_2\cdot
p_3)C_0(0,p_3-k,M_{\pi},M_K,M_K) \nonumber\\
&& -(M_K^2-M_{\pi}^2+4k\cdot p_3)C_0(-p_3,-k,0,M_K,M_K) \nonumber\\
&& +(M_K^2-M_{\pi}^2+4k\cdot p_3)(2M_K^2+2M_{\pi}^2+k\cdot
p_1+k\cdot p_2-2k\cdot p_3 \nonumber\\
&& \left. +2p_1\cdot p_2-p_1\cdot p_3-p_2\cdot
p_3)D_0(-p_3,-k,-p_3,m_{\gamma},M_{\pi},M_K,M_K)\right\} \,.
\end{eqnarray}

\subsubsection*{Figure~$7$}
We have
\begin{eqnarray}
\dot{\mathcal{Z}}_1^{(8)} &=&
-\frac{M_{\pi}^2}{M_K^2}\,\dot{\mathcal{Z}}_1^{(8')}\,=\,\frac{3}{2}\,\dot{\mathcal{Z}}_1^{(27)}
\\
&=& \frac{e^2}{6}\,M_{\pi}^2\left\{-2A_0(M_K) \right. \nonumber\\
&& +(5M_K^2+3M_{\pi}^2+k\cdot p_1+k\cdot p_2-2k\cdot p_3 \nonumber\\
&& +2p_1\cdot p_2-p_1\cdot p_3-p_2\cdot p_3)B_0(-k,0,M_K)
\nonumber\\
&& +(3M_K^2+5M_{\pi}^2+k\cdot p_1+k\cdot p_2-2k\cdot p_3 \nonumber\\
&& +2p_1\cdot p_2-p_1\cdot p_3-p_2\cdot p_3)B_0(-p_3,0,M_K)
\nonumber\\
&& -(M_K^2+3M_{\pi}^2+k\cdot p_1+k\cdot p_2-2k\cdot p_3 \nonumber\\
&& +2p_1\cdot p_2-p_1\cdot p_3-p_2\cdot p_3)B_0(p_3-k,M_K,M_K)
\nonumber\\
&& +(M_K^2-M_{\pi}^2+4k\cdot p_3)(M_K^2+3M_{\pi}^2+k\cdot p_1+k\cdot
p_2-2k\cdot p_3 \nonumber\\
&& \left. +2p_1\cdot p_2-p_1\cdot p_3-p_2\cdot
p_3)C_0(-p_3,-k,0,M_K,M_K)\right\} \,.
\end{eqnarray}

\subsubsection*{Figure~$8$}
We have
\begin{eqnarray}
\dot{\mathcal{Z}}_1^{(8)} &=&
\frac{3}{2}\,\dot{\mathcal{Z}}_1^{(27)} \\
&=& \frac{e^2}{6}\,(2M_K^2+2M_{\pi}^2+k\cdot p_1+k\cdot p_2-2k\cdot
p_3+2p_1\cdot
p_2-p_1\cdot p_3-p_2\cdot p_3)\times \nonumber\\
&& \left\{-2M_{\pi}^2(4\pi )^{-2}\left[ 1+\ln (m_{\gamma}^2)-\ln
(M_{\pi}M_K)\right]\right. \nonumber\\
&& +A_0(M_K)-2M_{\pi}^2B_0(-p_3,0,M_K)-M_K^2B_0(0,M_{\pi},M_K)
\nonumber\\
&& -2M_{\pi}^2B_1(-p_3,0,M_{\pi})-2M_{\pi}^2B_1(-k,0,M_K)
\nonumber\\
&& \left. +4M_{\pi}^2M_K^2C_0(-p_3,-p_3,0,M_{\pi},M_K)\right\}
\,, \\
\dot{\mathcal{Z}}_1^{(8')} &=&
-\frac{e^2}{6}\,M_K^2(2M_K^2+2M_{\pi}^2+k\cdot p_1+k\cdot
p_2-2k\cdot p_3+2p_1\cdot
p_2-p_1\cdot p_3-p_2\cdot p_3)\times \nonumber\\
&& \left\{-2(4\pi )^{-2}\left[ 1+\ln (m_{\gamma}^2)-\ln
(M_{\pi}M_K)\right]\right. \nonumber\\
&& +2B_0(-p_3,0,M_K)-B_0(0,M_{\pi},M_K)-2B_1(-p_3,0,M_{\pi})
\nonumber\\
&& \left.
-2B_1(-k,0,M_K)+4M_{\pi}^2C_0(-p_3,-p_3,0,M_{\pi},M_K)\right\} \,,
\\
\ddot{\mathcal{Z}}_1^{(8)} &=&
-\frac{M_{\pi}^2}{M_K^2}\,\ddot{\mathcal{Z}}_1^{(8')}\,=\,\frac{3}{2}\,\ddot{\mathcal{Z}}_1^{(27)}
\\
&=& -\frac{e^2}{6}\,M_{\pi}^2\,(2M_K^2+2M_{\pi}^2+k\cdot p_1+k\cdot
p_2-2k\cdot p_3+2p_1\cdot p_2-p_1\cdot p_3-p_2\cdot p_3)\times
\nonumber\\
&& \left[ -A_0(M_K)+2(M_K^2+M_{\pi}^2)B_0(-p_3,0,M_K)\right]\,.
\end{eqnarray}

\subsubsection*{Figure~$9$}
We have
\begin{eqnarray}
\dot{\mathcal{Z}}_1^{(8)} &=&
-\frac{M_{\pi}^2}{M_K^2}\,\dot{\mathcal{Z}}_1^{(8')}\,=\,-\dot{\mathcal{Z}}_1^{(27)}
\\
&=& \frac{e^2}{2}\,M_{\pi}^2\left\{-2A_0(M_K) \right. \nonumber\\
&& -2(k\cdot p_1+k\cdot p_2-2k\cdot p_3+2p_1\cdot p_2-p_1\cdot
p_3-p_2\cdot p_3)\times \nonumber\\
&& (4\pi )^{-2}\left[ 1+\ln (m_{\gamma}^2)-\ln (M_{\pi}M_K)\right]
\nonumber\\
&& +(8M_{\pi}^2+k\cdot p_1+k\cdot p_2-10k\cdot p_3 \nonumber\\
&& +2p_1\cdot p_2+3p_1\cdot p_3+3p_2\cdot p_3)B_0(-p_3,0,M_{\pi})
\nonumber\\
&& -2(k\cdot p_1+k\cdot p_2-2k\cdot p_3+2p_1\cdot p_2-p_1\cdot
p_3-p_2\cdot p_3)B_1(-p_3,0,M_{\pi}) \nonumber\\
&& +(8M_K^2-3k\cdot p_1-3k\cdot p_2-2k\cdot p_3 \nonumber\\
&& +2p_1\cdot p_2-p_1\cdot p_3-p_2\cdot p_3)B_0(-k,0,M_K)
\nonumber\\
&& -2(k\cdot p_1+k\cdot p_2-2k\cdot p_3+2p_1\cdot p_2-p_1\cdot
p_3-p_2\cdot p_3)B_1(-k,0,M_K) \nonumber\\
&& -(2M_{\pi}^2+k\cdot p_1+k\cdot p_2-4k\cdot p_3 \nonumber\\
&& +2p_1\cdot p_2-p_1\cdot p_3-p_2\cdot p_3)B_0(p_3-k,M_{\pi},M_K)
\nonumber\\
&& +2(-k\cdot p_1-k\cdot p_2+p_1\cdot p_3+p_2\cdot
p_3)B_1(p_3-k,M_{\pi},M_K) \nonumber\\
&& +4(k\cdot p_3)(k\cdot p_1+k\cdot p_2-2k\cdot p_3 \nonumber\\
&& +2p_1\cdot p_2-p_1\cdot p_3-p_2\cdot
p_3)C_0(-p_3,-k,m_{\gamma},M_{\pi},M_K) \nonumber\\
&& +8(k\cdot p_3)(p_1\cdot p_3+p_2\cdot
p_3)C_1(-p_3,-k,0,M_{\pi},M_K) \nonumber\\
&& \left. +8(k\cdot p_3)(k\cdot p_1+k\cdot
p_2)C_2(-p_3,-k,0,M_{\pi},M_K)\right\} \,.
\end{eqnarray}

\subsubsection*{Figure~$10$}
We have
\begin{equation}
\mathcal{T}_1^{(8)}\,=\,
\frac{3}{2}\,\mathcal{T}_1^{(27)}\,=\,-(M_K^2-M_{\pi}^2)^{-1}\dot{\mathcal{Z}}_1^{(8)}\,=\,-\frac{3}{2}\,(M_K^2-M_{\pi}^2)^{-1}\dot{\mathcal{Z}}_1^{(27)}
\,=\,0\,.
\end{equation}
This result is somewhat expected since the neutral pion is
\textit{electrically neutral}.

Our amplitude is ultraviolet divergent. So is the amplitude
calculated in Ref.~\cite{Bijnens:2004ku}. Only the sum of these is
ultraviolet finite. Our amplitude is also infrared divergent. We
shall show in the following section that the infrared divergence
cancels the one coming from real soft photon emission. The
cancelation occurs at the level of differential decay rates. Since
we are interested in evaluating the amplitude, we have to subtract
infrared divergence. The subtraction scheme is not unique. We will
adopt a \emph{minimal subtraction scheme} consisting on dropping out
only $\ln (m_{\gamma}^2)$ terms from the expression of the
amplitude.

\section{Infrared divergence}

The infrared divergence of the amplitude reads:
\begin{equation} \label{eq:bremsstrahlung-amplitude}
\mathcal{M}_{IR}\,=\,-\frac{e^2}{8\pi^2}\,\mathcal{M}_0\left[
1-(k\cdot p_3)\tau (-p_3,-k,M_{\pi},M_K)\right]\,\ln
(m_{\gamma}^2)\,.
\end{equation}
The divergence is canceled by the one generated from the emission of
a real soft photon. The energy $q$ of the latter being raised by the
detector resolution $\omega$. The cancelation takes place as usually
at the differential decay rate level. Let $\mathcal{M}_{\gamma}$
denotes the decay amplitude corresponding to the emission of one
real photon. We shall focus only on the infrared divergent part of
$\mathcal{M}_{\gamma}$. The Feynman diagrams one needs are drawn in
Fig.~$11$. The result is standard:
\begin{equation}
\mathcal{M}_{\gamma}\,=\,-e\mathcal{M}_0\left(\frac{k\cdot\varepsilon^*}{k\cdot
q}-\frac{p_3\cdot\varepsilon^*}{p_3\cdot q}\right)\,.
\end{equation}
One obtains after squaring the amplitude and summing over photon
polarizations
\begin{equation} \label{eq:bremsstrahlung-square}
\mid\mathcal{M}_{\gamma}\mid^2\,=\,-e^2\mid\mathcal{M}_0\mid^2\left[\frac{M_K^2}{(k\cdot
q)^2}-\frac{2(k\cdot p_3)}{(k\cdot q)(p_3\cdot
q)}+\frac{M_{\pi}^2}{(p_3\cdot q)^2}\right]\,.
\end{equation}
We shall now explicitly show the infrared divergence cancelation.
The infrared divergence coming from virtual photons is deduced from
Eqs.~(\ref{eq:rate}) and (\ref{eq:bremsstrahlung-amplitude})
\begin{equation}
d\Gamma_{IR}\,=\,-\frac{e^2}{4\pi^2}\,d\Gamma_0\left[ 1-(k\cdot
p_3)\tau (-p_3,-k,M_{\pi},M_K)\right]\,\ln (m_{\gamma}^2)\,.
\end{equation}
From Eq.~(\ref{eq:bremsstrahlung-square}), the differential decay
rate corresponding to the emission of one real photon reads in the
soft photon approximation
\begin{equation}
d\Gamma^{\gamma}\,=\,-e^2d\Gamma_0\int_{m_{\gamma}}^{\omega}\frac{d^3q}{(2\pi
)^32q}\left[\frac{M_K^2}{(k\cdot q)^2}-\frac{2(k\cdot p_3)}{(k\cdot
q)(p_3\cdot q)}+\frac{M_{\pi}^2}{(p_3\cdot q)^2}\right]\,.
\end{equation}
One can then easily check that
\begin{equation}
d\Gamma_{IR}\,+\,d\Gamma_{IR}^{\gamma}\,=\,0\,.
\end{equation}

\section{Results}

In this section we shall evaluate isospin breaking correction due to
explicit virtual photons. This correction depends only on $s_3$
thanks to the on-shell relation~(\ref{eq:onshell}). It is natural
then to study the variation of the correction with respect to $s_3$
in the physical region defined by Eq.~(\ref{eq:physical-region}). In
order to evaluate the size of the correction it is convenient to
compare it to the tree-level value or to the one-loop level
correction in the absence of isospin breaking. Since the latter
depends on two kinematical variables it is easier (but not less
instructive) to compare it to the former. This is done in
Fig.~($12$) where we plotted the Born amplitude squared
$\mid\mathcal{M}\mid^2$ (dashed curve) and the one-loop level
amplitude squared $\mid\mathcal{M}_0\mid^2+
2\,\Re\,(\mathcal{M}_0\mathcal{M}_1)$ (plain curve). We recall that
$\mathcal{M}_1$ stands for the explicit virtual photon correction.
In order to draw Fig.~($12$) we used the following numerical input:
\begin{eqnarray}
&& e^2\,=\,(4\pi )/(137.036)\,, \quad
M_{\pi}\,=\,0.134977\,\textrm{GeV}\,, \quad
M_K\,=\,0.495042\,\textrm{GeV}\,, \nonumber\\
&& C\,=\,-1.07\times 10^{-6}\,\textrm{GeV}^{-2}\,, \quad
G_8\,=\,5.45\,, \quad G_{27}\,=\,0.392\,.
\end{eqnarray}
Note also that the ultraviolet divergent terms ($\overline{\lambda}$
terms) have been dropped out and the scale $\mu$ was taken to be
$0.770\,\textrm{GeV}$. Let us comment the content of Fig.~($12$). To
this end, we shall write the amplitude squared as
\begin{equation}
\mid\mathcal{M}\mid^2\,=\,\mid\mathcal{M}_0\mid^2\left[ 1+\delta
(s_3)\right]\,.
\end{equation}
The ratio $\delta$ is given in Tab.~(\ref{tab:result}) for different
values of $s_3$. As can be seen from Tab.~(\ref{tab:result}), the
explicit virtual photon correction is about $5$ to $6\%$ the value
of the Born amplitude.

\section{Conclusion}

In this work we calculated isospin breaking corrections to the
process $K^+\rightarrow\pi^+\pi^0\pi^0$ due to virtual soft photons
at one-loop level and in the framework of chiral perturbation
theory. The corrections generated by $m_u-m_d$ and virtual hard
photons have been calculated in Ref.~\cite{Bijnens:2004ku}. The
corrections are individually ultraviolet divergent but jointly
ultraviolet finite. They are also infrared divergent. We showed that
this divergence is canceled at the differential decay level by the
one coming from real soft photons. We follow
Ref.~\cite{Bijnens:2004ku} and denote by $A_{00+}(0,0)$ the decay
amplitude for the process in question evaluated at the center of
Dalitz plot. Then the amplitude squared reads:
\begin{eqnarray}
\mid A_{00+}(0,0)\mid^2 &=& \underbrace{2.49\times
10^{-13}}_{\mathcal{O}(p^2)}+\underbrace{6.84\times
10^{-13}}_{\mathcal{O}(p^4)} \nonumber\\
&& +\underbrace{7.7\times
10^{-14}}_{\textrm{is.~br.~I}}+\underbrace{-1.3\times
10^{-14}}_{\textrm{is.~br.~II}}\,.
\end{eqnarray}
In the first line of the equation we reported the value of the
amplitude squared at leading and next-to-leading chiral orders and
in the absence of isospin breaking~\cite{Bijnens:2002vr}. The first
term in the second line represents isospin breaking corrections due
to $m_u-m_d$ and to hard virtual photons~\cite{Bijnens:2004ku}. The
second term has been calculated in the present work and corresponds
to the isospin breaking correction due to soft virtual photons.
Adding all this together, isospin breaking correction to the
amplitude squared represents $7\%$ its one-loop value in the absence
of isospin breaking and at the center of Dalitz plot.

\appendix

\section{Loop integrals}

We use dimensional regularization and adopt the
$\overline{\textrm{MS}}$ subtraction scheme
\begin{equation}
\overline{\lambda}\,\doteq\,-\frac{1}{32\pi^2}\left[\frac{2}{4-n}+1-\gamma+\ln
(4\pi )\right]\,, \end{equation} where $n$ is space-time dimension
and $\gamma$ the Euler constant. All the technical material
necessary for the calculation of one-loop integrals is given in the
appendix of Ref.~\cite{Nehme:2003bz}.

It is convenient to take the following notations:
\begin{eqnarray}
\sigma_{P} &\doteq& \sqrt{1-\frac{4M_P^2}{s_3}}\,,  \\
\sigma_{PP} &\doteq& \frac{\sigma_P-1}{\sigma_P+1}\,, \\
\sigma_{K\pi} &\doteq&
\frac{\sqrt{(M_K+M_{\pi})^2-s_3}-\sqrt{(M_K-M_{\pi})^2-s_3}}{\sqrt{(M_K+M_{\pi})^2-s_3}+\sqrt{(M_K-M_{\pi})^2-s_3}}\,,
\\
\lambda^{1/2}(s_3,M_{\pi}^2,M_K^2) &\doteq&
\sqrt{(M_K+M_{\pi})^2-s_3}\sqrt{(M_K-M_{\pi})^2-s_3}\,.
\end{eqnarray}

\subsection*{$A$ integrals}

The one-point function reads:
\begin{equation}
A_0(m)\,=\,m^2\left[
-2\overline{\lambda}-\frac{1}{16\pi^2}\,\ln\left(\frac{m^2}{\mu^2}\right)\right]\,,
\end{equation}
where $\mu$ an arbitrary scale with mass dimension.

\subsection*{$B$ integrals}

We need the following two-point functions:
\begin{eqnarray}
B_0(-p_3,0,M_{\pi}) &=&
\frac{A_0(M_{\pi})}{M_{\pi}^2}+\frac{1}{16\pi^2}\,, \\
B_1(-p_3,0,M_{\pi}) &=&
-\frac{1}{2}\,\frac{A_0(M_{\pi})}{M_{\pi}^2}\,, \\
B_0(-k,0,M_K) &=& \frac{A_0(M_K)}{M_K^2}+\frac{1}{16\pi^2}\,, \\
B_1(-k,0,M_K) &=& -\frac{1}{2}\,\frac{A_0(M_K)}{M_K^2}\,, \\
B_0(0,M_{\pi},M_K) &=&
\frac{A_0(M_{\pi})}{M_{\pi}^2}+\frac{1}{16\pi^2}\,\frac{M_K^2}{M_K^2-M_{\pi}^2}\,\ln\left(\frac{M_{\pi}^2}{M_K^2}\right)\,,
\\
B_0(-p_3,0,M_K) &=& \frac{A_0(M_K)}{M_K^2}+\frac{1}{16\pi^2}\left[
1-\left( 1-\frac{M_K^2}{M_{\pi}^2}\right)\ln\left(
1-\frac{M_{\pi}^2}{M_K^2}\right)\right]\,, \\
\Re\,B_0(-k,0,M_{\pi}) &=&
\frac{A_0(M_{\pi})}{M_{\pi}^2}+\frac{1}{16\pi^2}\left[ 1-\left(
1-\frac{M_{\pi}^2}{M_K^2}\right)\ln\left(
\frac{M_K^2}{M_{\pi}^2}-1\right)\right]\,, \\
\Im\,B_0(-k,0,M_{\pi}) &=& \frac{1}{16\pi}\left(
1-\frac{M_{\pi}^2}{M_K^2}\right)\,, \\
\Re\,B_0(k-p_3,M_{\pi},M_{\pi}) &=&
\frac{A_0(M_{\pi})}{M_{\pi}^2}+\frac{1}{16\pi^2}\left[
1+\sigma_{\pi}\ln (-\sigma_{\pi\pi})\right]\,, \\
\Im\,B_0(k-p_3,M_{\pi},M_{\pi}) &=& \frac{\sigma_{\pi}}{16\pi}\,, \\
B_0(p_3-k,M_K,M_K) &=& \frac{A_0(M_K)}{M_K^2} \nonumber\\
&& +\frac{1}{16\pi^2}\left[
1-2\left(\frac{4M_K^2}{s_3}-1\right)^{1/2}\textrm{arctan}\,\sqrt{\frac{s_3}{4M_K^2-s_3}}\right]\,,
\\
B_0(p_3-k,M_{\pi},M_K) &=&
\frac{1}{2}\,\frac{A_0(M_K)}{M_K^2}+\frac{1}{2}\,\frac{A_0(M_{\pi})}{M_{\pi}^2}
\nonumber\\
&& +\frac{1}{16\pi^2}\left[
1+\frac{M_K^2-M_{\pi}^2}{s_3}\,\ln\left(\frac{M_{\pi}}{M_K}\right)\right.
\nonumber\\
&& \left. -\frac{1}{s_3}\,\lambda^{1/2}(s_3,M_{\pi}^2,M_K^2)\ln
(\sigma_{K\pi})\right]\,, \\
B_1(p_3-k,M_{\pi},M_K) &=& \frac{1}{2s_3}\left[
A_0(M_{\pi})-A_0(M_K)\right. \nonumber\\
&& \left. -(s_3-M_K^2+M_{\pi}^2)B_0(p_3-k,M_{\pi},M_K)\right]\,.
\end{eqnarray}

\subsection*{$\tau$ integrals}

The definition and expression of $\tau$ integrals can be found in
Ref.~\cite{Nehme:2003bz}. Note the particular expression
\begin{equation}
\tau
(-p_3,-k,M_{\pi},M_K)\,=\,-2\lambda^{-1/2}(s_3,M_{\pi}^2,M_K^2)\ln
(\sigma_{K\pi})\,.
\end{equation}

\subsection*{$C$ integrals}

The three-point functions needed for our purposes are:
\begin{eqnarray}
&& C_0(-p_3,-p_3,0,M_{\pi},M_K)\,=\, \nonumber\\
&& \frac{1}{16\pi^2}\left[\frac{1}{M_{\pi}^2}\,\ln\left(
1-\frac{M_{\pi}^2}{M_K^2}\right)
+\frac{1}{M_K^2-M_{\pi}^2}\,\ln\left(\frac{M_{\pi}^2}{M_K^2}\right)\right]\,,
\\
&& \Re\,C_0(-k,-k,0,M_{\pi},M_K)\,=\, \nonumber\\
&&
-\frac{1}{16\pi^2M_K^2}\left[\ln\left(\frac{M_{\pi}^2}{M_K^2-M_{\pi}^2}\right)
-\frac{M_K^2}{M_K^2-M_{\pi}^2}\,\ln\left(\frac{M_{\pi}^2}{M_K^2}\right)\right]\,,
\\
&& \Im\,C_0(-k,-k,0,M_{\pi},M_K)\,=\,-\frac{1}{16\pi M_K^2}\,, \\
&& C_0(0,p_3-k,M_{\pi},M_K,M_K)\,=\, \nonumber\\
&& \frac{1}{16\pi^2}\,\frac{1}{M_K^2-M_{\pi}^2}\left[\left(
1-\frac{M_K^2-M_{\pi}^2}{s_3}\right)\ln\left(\frac{M_{\pi}}{M_K}\right)\right.
\nonumber\\
&& \left.
-2\sqrt{\frac{4M_K^2}{s_3}-1}\,\textrm{arctan}\,\sqrt{\frac{s_3}{4M_K^2-s_3}}+\frac{1}{s_3}\,\lambda^{1/2}(s_3,M_{\pi}^2,M_K^2)\ln
(\sigma_{K\pi})\right]\,, \\
&& \Re\,C_0(0,k-p_3,M_K,M_{\pi},M_{\pi})\,=\, \nonumber\\
&& \frac{1}{16\pi^2}\,\frac{1}{M_K^2-M_{\pi}^2}\left[\left(
1+\frac{M_K^2-M_{\pi}^2}{s_3}\right)\ln\left(\frac{M_{\pi}}{M_K}\right)
\right. \nonumber\\
&& \left. -\sigma_{\pi}\ln
(-\sigma_{\pi\pi})-\frac{1}{s_3}\,\lambda^{1/2}(s_3,M_{\pi}^2,M_K^2)\ln
(\sigma_{K\pi})\right]\,, \\
&&
\Im\,C_0(0,k-p_3,M_K,M_{\pi},M_{\pi})\,=\,-\frac{1}{16\pi}\,\frac{\sigma_{\pi}}{M_K^2-M_{\pi}^2}\,,
\\
&&
C_0(-p_3,-k,0,M_K,M_K)\,=\,\frac{1}{16\pi^2}\,\frac{1}{M_KM_{\pi}}\,\frac{\sigma_{K\pi}}{1-\sigma_{K\pi}^2}\times
\nonumber\\
&& \left\{
-\ln^2(\sigma_{KK})-\frac{1}{2}\,\ln^2(\sigma_{K\pi})-\frac{1}{2}\,\ln^2\left(\frac{M_{\pi}}{M_K}\right)
\right. \nonumber\\
&& +\ln
(\sigma_{K\pi})\left[\ln\left(\frac{M_KM_{\pi}}{M_K^2-M_{\pi}^2}\right)+\ln\left(\frac{M_K^2}{M_K^2-M_{\pi}^2}\right)\right]
\nonumber\\
&& +\textrm{Li}_2\left(
1-\frac{M_{\pi}}{M_K}\,\sigma_{K\pi}\right)+\textrm{Li}_2\left(
1-\frac{M_K}{M_{\pi}}\,\sigma_{K\pi}\right) \nonumber\\
&& -\textrm{Li}_2\left(
1-\frac{M_K}{M_{\pi}}\,\frac{\sigma_{K\pi}}{\sigma_{KK}}\right)-\textrm{Li}_2\left(
1-\frac{M_K}{M_{\pi}}\,\sigma_{KK}\sigma_{K\pi}\right) \nonumber\\
&& \left. -\textrm{Li}_2\left(
1-\frac{M_{\pi}}{M_K}\,\frac{\sigma_{K\pi}}{\sigma_{KK}}\right)-\textrm{Li}_2\left(
1-\frac{M_{\pi}}{M_K}\,\sigma_{KK}\sigma_{K\pi}\right)\right\}\,, \\
&&
\Re\,C_0(-p_3,-k,0,M_{\pi},M_{\pi})\,=\,\frac{1}{16\pi^2}\,\frac{1}{M_KM_{\pi}}\,\frac{\sigma_{K\pi}}{1-\sigma_{K\pi}^2}\times
\nonumber\\
&& \left\{
-\frac{\pi^2}{3}-\frac{1}{2}\ln^2(\sigma_{K\pi})-\ln^2(-\sigma_{\pi\pi})-\frac{1}{2}\,\ln^2\left(\frac{M_{\pi}}{M_K}\right)
\right. \nonumber\\
&& +\ln
(\sigma_{K\pi})\left[\ln\left(\frac{M_{\pi}^2}{M_K^2-M_{\pi}^2}\right)
+\ln\left(\frac{M_KM_{\pi}}{M_K^2-M_{\pi}^2}\right)\right]
\nonumber\\
&& +\frac{1}{2}\ln^2\left(
1-\frac{M_K}{M_{\pi}}\,\frac{\sigma_{K\pi}}{\sigma_{\pi\pi}}\right)+\frac{1}{2}\ln^2\left(
1-\frac{M_K}{M_{\pi}}\,\sigma_{K\pi}\sigma_{\pi\pi}\right)
\nonumber\\
&& +\frac{1}{2}\ln^2\left(
1-\frac{M_{\pi}}{M_K}\,\frac{\sigma_{K\pi}}{\sigma_{\pi\pi}}\right)+\frac{1}{2}\ln^2\left(
1-\frac{M_{\pi}}{M_K}\,\sigma_{K\pi}\sigma_{\pi\pi}\right)
\nonumber\\
&& +\textrm{Li}_2\left( 1-\frac{M_{\pi}}{M_K}\,\sigma_{K\pi}\right)
+\textrm{Li}_2\left( 1-\frac{M_K}{M_{\pi}}\,\sigma_{K\pi}\right)
\nonumber\\
&&
+\textrm{Li}_2\left(\frac{M_{\pi}\sigma_{\pi\pi}}{M_{\pi}\sigma_{\pi\pi}-M_K\sigma_{K\pi}}\right)
+\textrm{Li}_2\left(\frac{M_{\pi}}{M_{\pi}-M_K\sigma_{K\pi}\sigma_{\pi\pi}}\right)
\nonumber\\
&& \left.
+\textrm{Li}_2\left(\frac{M_K\sigma_{\pi\pi}}{M_K\sigma_{\pi\pi}-M_{\pi}\sigma_{K\pi}}\right)
+\textrm{Li}_2\left(\frac{M_K}{M_K-M_{\pi}\sigma_{K\pi}\sigma_{\pi\pi}}\right)
\right\}\,, \\
&&
\Im\,C_0(-p_3,-k,0,M_{\pi},M_{\pi})\,=\,-\frac{1}{16\pi}\,\frac{1}{M_KM_{\pi}}\,\frac{\sigma_{K\pi}}{1-\sigma_{K\pi}^2}\times
\nonumber\\
&& \left[ -2\ln (\sigma_{K\pi})+2\ln (-\sigma_{\pi\pi})+\ln\left(
1-\frac{M_K}{M_{\pi}}\,\frac{\sigma_{K\pi}}{\sigma_{\pi\pi}}\right)
\right. \nonumber\\
&& \left. -\ln\left(
1-\frac{M_K}{M_{\pi}}\,\sigma_{K\pi}\sigma_{\pi\pi}\right)
+\ln\left(
1-\frac{M_{\pi}}{M_K}\,\frac{\sigma_{K\pi}}{\sigma_{\pi\pi}}\right)-\ln\left(
1-\frac{M_{\pi}}{M_K}\,\sigma_{K\pi}\sigma_{\pi\pi}\right)
\right]\,,
\\
&&
C_0(-p_3,-k,m_{\gamma},M_{\pi},M_K)\,=\,\frac{1}{16\pi^2}\,\frac{1}{M_KM_{\pi}}\,\frac{\sigma_{K\pi}}{1-\sigma_{K\pi}^2}\times
\nonumber\\
&& \left\{\ln (\sigma_{K\pi})\left[ 2\ln
(1-\sigma_{K\pi}^2)-\frac{1}{2}\,\ln
(\sigma_{K\pi})-\ln\left(\frac{m_{\gamma}^2}{M_KM_{\pi}}\right)\right]\right.
\nonumber\\
&&
-\frac{\pi^2}{6}+\frac{1}{2}\,\ln^2\left(\frac{M_{\pi}}{M_K}\right)
+\textrm{Li}_2(\sigma_{K\pi}^2) \nonumber\\
&& \left. +\textrm{Li}_2\left(
1-\frac{M_{\pi}}{M_K}\,\sigma_{K\pi}\right) +\textrm{Li}_2\left(
1-\frac{M_K}{M_{\pi}}\,\sigma_{K\pi}\right)\right\}\,, \\
&& C_1(-p_3,-k,0,M_{\pi},M_K)\,=\, \nonumber\\
&& \lambda^{-1}(s_3,M_{\pi}^2,M_K^2)\left[
(M_K^2+M_{\pi}^2-s_3)B_0(-p_3,0,M_{\pi}) \right. \nonumber\\
&& \left.
-2M_K^2B_0(-k,0,M_K)+(M_K^2-M_{\pi}^2+s_3)B_0(p_3-k,M_{\pi},M_K)\right]\,,
\\
&& C_2(-p_3,-k,0,M_{\pi},M_K)\,=\, \nonumber\\
&& \lambda^{-1}(s_3,M_{\pi}^2,M_K^2)\left[
-2M_{\pi}^2B_0(-p_3,0,M_{\pi}) \right. \nonumber\\
&& \left.
+(M_K^2+M_{\pi}^2-s_3)B_0(-k,0,M_K)-(M_K^2-M_{\pi}^2-s_3)B_0(p_3-k,M_{\pi},M_K)\right]\,.
\end{eqnarray}

\subsection*{$D$ integrals}

We have one complex and one real four-point function in the physical
region. The complex one reads:
\begin{eqnarray}
&& \Re\,D_0(-k,-p_3,-k,m_{\gamma},M_K,M_{\pi},M_{\pi})\,=\,
\nonumber\\
&&
\frac{1}{16\pi^2}\,\frac{1}{M_KM_{\pi}}\,\frac{1}{M_K^2-M_{\pi}^2}\,\frac{\sigma_{K\pi}}{1-\sigma_{K\pi}^2}\times
\nonumber\\
&& \left\{\frac{\pi^2}{6}+2\ln (\sigma_{K\pi})\left[\ln
(1-\sigma_{K\pi}^2)-\ln\left(\frac{M_{\pi}m_{\gamma}}{M_K^2-M_{\pi}^2}\right)\right]\right.
\nonumber\\
&& -\frac{1}{2}\,\ln^2\left(
1-\frac{M_K}{M_{\pi}}\,\frac{\sigma_{K\pi}}{\sigma_{\pi\pi}}\right)
-\frac{1}{2}\,\ln^2\left(
1-\frac{M_K}{M_{\pi}}\,\sigma_{K\pi}\sigma_{\pi\pi}\right)
\nonumber\\
&& -\frac{1}{2}\,\ln^2\left(
1-\frac{M_{\pi}}{M_K}\,\frac{\sigma_{K\pi}}{\sigma_{\pi\pi}}\right)
-\frac{1}{2}\,\ln^2\left(
1-\frac{M_{\pi}}{M_K}\,\sigma_{K\pi}\sigma_{\pi\pi}\right)
\nonumber\\
&& +\ln^2\left(\frac{M_{\pi}}{M_K}\right)
+\ln^2(-\sigma_{\pi\pi})+\textrm{Li}_2(\sigma_{K\pi}^2) \nonumber\\
&&
-\textrm{Li}_2\left(\frac{M_{\pi}\sigma_{\pi\pi}}{M_{\pi}\sigma_{\pi\pi}-M_K\sigma_{K\pi}}\right)
-\textrm{Li}_2\left(\frac{M_{\pi}}{M_{\pi}-M_K\sigma_{K\pi}\sigma_{\pi\pi}}\right)
\nonumber\\
&& \left.
-\textrm{Li}_2\left(\frac{M_K\sigma_{\pi\pi}}{M_K\sigma_{\pi\pi}-M_{\pi}\sigma_{K\pi}}\right)
-\textrm{Li}_2\left(\frac{M_K}{M_K-M_{\pi}\sigma_{K\pi}\sigma_{\pi\pi}}\right)\right\}\,,
\\
&& \Im\,D_0(-k,-p_3,-k,m_{\gamma},M_K,M_{\pi},M_{\pi})\,=\,
\nonumber\\
&&
\frac{1}{16\pi}\,\frac{1}{M_KM_{\pi}}\,\frac{1}{M_K^2-M_{\pi}^2}\,\frac{\sigma_{K\pi}}{1-\sigma_{K\pi}^2}\times
\nonumber\\
&& \left\{-2\ln (\sigma_{K\pi})+2\ln (-\sigma_{\pi\pi})\right.
\nonumber\\
&& +\ln\left(
1-\frac{M_K}{M_{\pi}}\,\frac{\sigma_{K\pi}}{\sigma_{\pi\pi}}\right)
-\ln\left(
1-\frac{M_K}{M_{\pi}}\,\sigma_{K\pi}\sigma_{\pi\pi}\right)
\nonumber\\
&& \left. +\ln\left(
1-\frac{M_{\pi}}{M_K}\,\frac{\sigma_{K\pi}}{\sigma_{\pi\pi}}\right)
-\ln\left(
1-\frac{M_{\pi}}{M_K}\,\sigma_{K\pi}\sigma_{\pi\pi}\right)\right\}\,.
\end{eqnarray}
The real one is given by:
\begin{eqnarray}
&& D_0(-p_3,-k,-p_3,m_{\gamma},M_{\pi},M_K,M_K)\,=\,
\nonumber\\
&&
-\frac{1}{16\pi^2}\,\frac{1}{M_KM_{\pi}}\,\frac{1}{M_K^2-M_{\pi}^2}\,\frac{\sigma_{K\pi}}{1-\sigma_{K\pi}^2}\times
\nonumber\\
&& \left\{-\frac{\pi^2}{6}+2\ln (\sigma_{K\pi})\left[\ln
(1-\sigma_{K\pi}^2)-\ln\left(\frac{M_Km_{\gamma}}{M_K^2-M_{\pi}^2}\right)\right]\right.
\nonumber\\
&& +\ln^2\left(\frac{M_{\pi}}{M_K}\right)
+\ln^2(\sigma_{KK})+\textrm{Li}_2(\sigma_{K\pi}^2) \nonumber\\
&& +\textrm{Li}_2\left(
1-\frac{M_K}{M_{\pi}}\,\frac{\sigma_{K\pi}}{\sigma_{KK}}\right)
+\textrm{Li}_2\left(
1-\frac{M_K}{M_{\pi}}\,\sigma_{KK}\sigma_{K\pi}\right)
\nonumber\\
&& \left. +\textrm{Li}_2\left(
1-\frac{M_{\pi}}{M_K}\,\frac{\sigma_{K\pi}}{\sigma_{KK}}\right)
+\textrm{Li}_2\left(
1-\frac{M_{\pi}}{M_K}\,\sigma_{KK}\sigma_{K\pi}\right)\right\} \,.
\end{eqnarray}

\pagebreak

\begin{table}
\begin{center}
\begin{tabular}{|c||c|c|c|c|c|c|c|}
\hline $r$ & $1$ & $1.12$ & $1.24$ &
$s_0/(4M_{\pi}^2)$ & $1.48$ & $1.60$ & $(M_K-M_{\pi})^2/(4M_{\pi}^2)$ \\
\hline $10^2\,\delta$ & $-5.78$ & $-5.59$ & $-5.43$ & $-5.26$ &
$-5.13$ & $-4.99$ & $-5.44$ \\
\hline
\end{tabular}
\end{center}
\caption{\label{tab:result} The size of explicit virtual photon
correction compared to the Born amplitude. We used the notations $r$
and $s_0$ for the ratio $s/(4M_{\pi}^2)$ and the Dalitz plot center
$(s_1+s_2+s_3=M_K^2+3M_{\pi}^2)/3$, respectively.}
\end{table}

\end{document}